\pdfoutput=1 %
\newif\ifpreprint \preprinttrue
\newif\ifanon     \anonfalse
\ifanon\PassOptionsToClass{anonymous}{acmart}\fi
\ifpreprint\PassOptionsToClass{nonacm}{acmart}\else\PassOptionsToClass{review}{acmart}\fi
\documentclass[sigconf]{acmart}

\usepackage{booktabs}
\usepackage{tabularx}
\usepackage{multirow}
\usepackage{tikz}
\usetikzlibrary{arrows.meta,positioning,fit,backgrounds,calc,decorations.pathreplacing}
\usepackage{xcolor}
\allowdisplaybreaks[2]

\DeclareRobustCommand{\vtag}[2]{\mbox{#1\if\relax\detokenize{#2}\relax\else-#2\fi}}
\DeclareRobustCommand{\dna}[1][]{\vtag{DNA}{#1}}
\DeclareRobustCommand{\ro}[1][]{\vtag{CrossAttn}{#1}}
\DeclareRobustCommand{\dnat}[1][]{\vtag{DNA-T}{#1}}
\newcommand{\pp}{\,pt}
\newcommand{\rblk}{\mathcal{A}}
\newcommand{\wblk}{\mathcal{B}}
\newcommand{\R}{\mathbb{R}}
\newcommand{\Sb}{\mathbf{S}}
\newcommand{\Ub}{\mathbf{U}}
\newcommand{\Fb}{\mathbf{F}}
\newcommand{\Qb}{\mathbf{Q}}
\newcommand{\Kb}{\mathbf{K}}
\newcommand{\Vb}{\mathbf{V}}
\newcommand{\Pb}{\mathbf{P}}

\ifpreprint\else
\setcopyright{acmlicensed}
\copyrightyear{2027}
\acmYear{2027}
\acmDOI{XXXXXXX.XXXXXXX}
\acmConference[KDD '27]{the 33rd ACM SIGKDD Conference on Knowledge Discovery and Data Mining}{August 1--5, 2027}{TBD}
\acmISBN{978-1-4503-XXXX-X/2027/08}
\fi
\begin{document}

\title{DNA: Unlocking Depth Scaling in CTR Models}

\ifpreprint\settopmatter{authorsperrow=3}\else\settopmatter{authorsperrow=4}\fi
\author{Caiyuan Li}
\affiliation{%
  \institution{Alibaba Group}
  \city{Hangzhou}
  \country{China}}
\email{licaiyuan.lcy@alibaba-inc.com}

\author{Wenjie Wu}
\affiliation{%
  \institution{Alibaba Group}
  \city{Hangzhou}
  \country{China}}
\email{wuwenjie.wwj@alibaba-inc.com}

\author{Yuheng Huang}
\affiliation{%
  \institution{Alibaba Group}
  \city{Hangzhou}
  \country{China}}
\email{hexing.hyh@alibaba-inc.com}

\author{Xiao Zhou}
\affiliation{%
  \institution{Alibaba Group}
  \city{Hangzhou}
  \country{China}}
\email{xiaoman.zx@taobao.com}

\author{Silu Zhou}
\affiliation{%
  \institution{Alibaba Group}
  \city{Hangzhou}
  \country{China}}
\email{silu.zsl@alibaba-inc.com}

\author{Ke Chen}
\affiliation{%
  \institution{Alibaba Group}
  \city{Hangzhou}
  \country{China}}
\email{ck325090@alibaba-inc.com}

\author{Zinan Zhao}
\affiliation{%
  \institution{Alibaba Group}
  \city{Hangzhou}
  \country{China}}
\email{zhaozinan.zzn@alibaba-inc.com}

\author{Xinrui Zeng}
\affiliation{%
  \institution{Alibaba Group}
  \city{Hangzhou}
  \country{China}}
\email{zengxinrui.zxr@alibaba-inc.com}

\author{Hanmo Xu}
\affiliation{%
  \institution{Alibaba Group}
  \city{Hangzhou}
  \country{China}}
\email{xuhanmo.xhm@alibaba-inc.com}

\author{Chaoqun Song}
\affiliation{%
  \institution{Alibaba Group}
  \city{Hangzhou}
  \country{China}}
\email{scq496778@alibaba-inc.com}

\author{Hua Zong}
\affiliation{%
  \institution{Alibaba Group}
  \city{Beijing}
  \country{China}}
\email{yuhuan.zh@alibaba-inc.com}

\author{Zongyuan Wu}
\affiliation{%
  \institution{Alibaba Group}
  \city{Hangzhou}
  \country{China}}
\email{zongyuan.wuzy@alibaba-inc.com}

\author{Hongbin Zhou}
\affiliation{%
  \institution{Alibaba Group}
  \city{Hangzhou}
  \country{China}}
\email{hongbin.zhouhb@alibaba-inc.com}

\author{Huimin Yi}
\affiliation{%
  \institution{Alibaba Group}
  \city{Beijing}
  \country{China}}
\email{huimin.yhm@alibaba-inc.com}

\author{Zhixuan Zhang}
\affiliation{%
  \institution{Alibaba Group}
  \city{Hangzhou}
  \country{China}}
\email{zhibing.zzx@taobao.com}

\author{Dan Ou}
\affiliation{%
  \institution{Alibaba Group}
  \city{Hangzhou}
  \country{China}}
\email{oudan.od@taobao.com}

\author{Haihong Tang}
\affiliation{%
  \institution{Alibaba Group}
  \city{Hangzhou}
  \country{China}}
\email{piaoxue@taobao.com}

\renewcommand{\shortauthors}{Li et al.}

\begin{abstract}
Long user behavior sequences are among the strongest signals for click-through rate (CTR) prediction. Cross-attention makes these sequences practical for industrial ranking by keeping compute linear in sequence length. Yet the gains from stacking cross-attention layers taper off: on Taobao search, growing a cross-attention stack from 16 to 28 layers adds only 0.02 percentage points (pt) of AUC. We argue that the key--value side is the bottleneck. Each layer projects the same behavior sequence into keys and values, so adding depth gives the model more reads of a fixed sequence. We propose \dna{} (Dual-stream Nexus Attention), which lets the sequence itself evolve with depth. At each layer, the request representation (the search query, user features and scene) first reads the sequence. Each behavior token then attends to the updated request representation and uses what it reads to update itself, so the next layer reads a sequence shaped by the current request. The update runs once per request, all candidates share the resulting sequence, and both the reads and the update keep linear complexity in sequence length. At comparable forward FLOPs, \dna{} improves AUC over cross-attention stacks by 0.08 to 0.26\pp{} and keeps improving in the depth range where those stacks level off. \dna[4] is deployed in Taobao search. In an online A/B test against the production model, it improves CTR by 0.8\% and GMV by 0.5\%.

\end{abstract}



\maketitle

\section{Introduction}

\begin{figure}[t]
  \centering
  \includegraphics[width=\columnwidth]{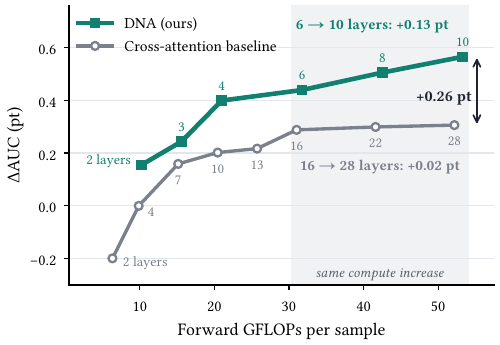}
  \Description{AUC gain versus forward GFLOPs per sample for DNA and the cross-attention baseline. Each marker is labeled with its number of layers. Over the same compute increase, the baseline grows from 16 to 28 layers and gains 0.02 points, while DNA grows from 6 to 10 layers and gains 0.13 points, ending 0.26 points above the baseline.}
  \caption{Depth scaling on Taobao search. Over the same compute increase (shaded), growing the cross-attention baseline from 16 to 28 layers adds 0.02\pp{} of AUC, while growing \dna{} from 6 to 10 layers adds 0.13\pp{}. Each marker is one model, labeled with its number of layers. $\Delta$AUC is relative to the four-layer baseline, \ro[4].}
  \label{fig:teaser}
\end{figure}

\begin{figure*}[t]
  \centering
  \includegraphics[width=\textwidth]{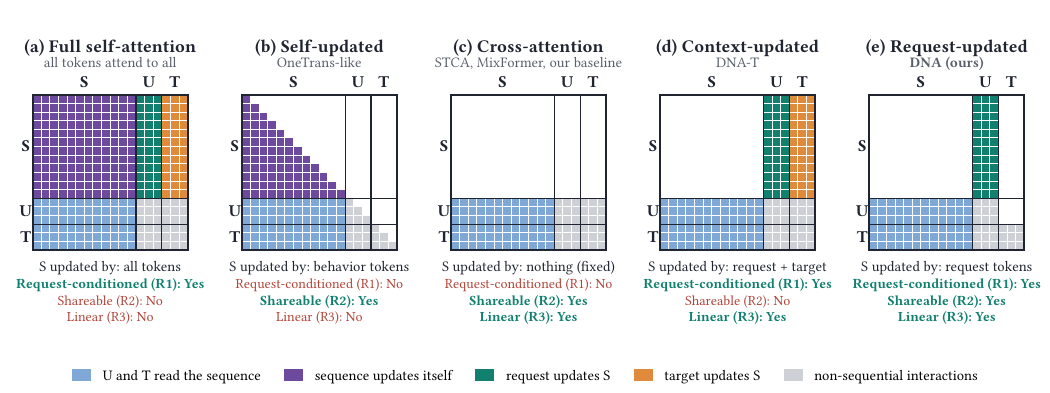}
  \Description{Five schematic panels compare attention involving the behavior sequence and interactions among non-sequential features. Each panel reports whether the sequence incorporates the current request, can be shared across candidates, and has linear attention cost.}
  \caption{Schematic information flow in long-sequence ranking. Rows receive information from columns. $\Sb$, $\Ub$ and $\mathbf{T}$ denote the behavior sequence, request features and features specific to one candidate. The full non-sequential representation $\Fb$ uses both request and candidate features. Blue, purple, teal and orange mark attention involving $\Sb$, and gray marks interactions among non-sequential features. The same colors are used in all figures: teal for the request stream and \dna{}, orange for candidate features and \dnat{}. R1 asks whether the sequence incorporates features of the current request. R2 asks whether candidates within a request can share the same sequence state. R3 asks whether attention cost grows linearly with sequence length $L$ at fixed non-sequential token counts and model width.}
  \label{fig:maps}
\end{figure*}

A search ranker must use a long record of user behavior to score many candidates within a tight latency budget. Cross-attention is a common way to do this. Representations of the request and candidate read the behavior sequence, which supplies keys and values~\cite{zhou2018din,pi2020sim,chang2023twin,guan2025stca,huang2026mixformer,chai2025longer}. With a fixed number of queries, its cost grows linearly with sequence length. Candidate-independent keys and values can also be reused across all candidates in a request~\cite{guan2025stca,guo2025roo,zhang2025onetrans}. Industrial results show why this design is attractive. LONGER retains over 95\% of the improvement of using every behavior as a query when it samples 40\% of them~\cite{chai2025longer}. STCA models 10,000 behaviors with 21 GFLOPs, compared with 156 GFLOPs for a self-attention Transformer over 8,000 behaviors in its study~\cite{guan2025stca}.

Deployed cross-attention models typically remain shallow. DIN, SIM and TWIN use one target-attention layer~\cite{zhou2018din,pi2020sim,chang2023twin,lv2024marm}. STCA and MixFormer use four~\cite{guan2025stca,huang2026mixformer}, and LONGER reads the full sequence only in its first layer~\cite{chai2025longer}. Where deeper stacks have been tested, their gains taper off. LONGER reports marginal gains beyond a certain depth~\cite{chai2025longer}. In ULTRA-HSTU, target cross-attention reduces normalized entropy by 0.14\% at both nine and twelve layers relative to three layers. Its twelve-layer self-attention model reduces it by 0.46\%~\cite{ding2026ultrahstu}. We observe the same pattern on Taobao search: growing a cross-attention stack from 16 to 28 layers adds only 0.02 percentage points of AUC (Figure~\ref{fig:teaser}).

We argue that this pattern points to a bottleneck on the key--value side of cross-attention. Each added layer refines the queries, that is, how the candidate reads the sequence, but every layer reads the same behaviors. In STCA and MixFormer, for example, each layer learns its own projection of the original behavior tokens~\cite{guan2025stca,huang2026mixformer}. A behavior token never receives information from the request, the candidate or other behaviors, however deep the stack, so added depth buys more reads of a fixed sequence. To get more from depth, the sequence itself has to evolve from layer to layer. Our ablations support this reading: an update that transforms each behavior on its own, without attending to other tokens, is far weaker than one that lets the behavior take in context (Section~\ref{sec:who}).

How should the sequence be updated? A ranking model imposes three requirements. The updated sequence should reflect the current request (R1), so that the same history can be read differently for different searches. It should be shareable by all candidates of a request (R2), so that scoring thousands of candidates does not multiply its cost. Its attention cost should grow linearly with sequence length (R3). Figure~\ref{fig:maps} checks five designs against these requirements.

Most ranking models that update the sequence take information for each behavior from other behaviors (Figure~\ref{fig:maps}b). HSTU and OneTrans use causal self-attention across behavior tokens~\cite{zhai2024hstu,zhang2025onetrans}. EST aggregates the behaviors most similar in content to each token~\cite{liu2026est}. This route has three costs. First, the update stays inside the sequence and misses R1. In OneTrans, for example, the non-sequential tokens follow the behaviors under the causal mask, so a past click on running shoes is encoded the same way whether the user now searches for ``track pants'' or ``water bottle''. Second, full causal self-attention over $L$ behaviors has quadratic cost in $L$, which misses R3. HSTU and OneTrans contain that cost with techniques such as sparsification, pruning and caching. Third, a single causal attention stack needs one order for all the groups it processes. Clicks, purchases and exposures over different time windows can be interleaved by time or concatenated in a chosen order with separators. OneTrans reports a 0.09\% AUC difference between two ordering rules~\cite{zhang2025onetrans}. Such heterogeneous groups are common in industrial ranking systems and can be difficult to merge. HyFormer avoids the merge by giving each group its own query tokens, with keys and values drawn only from the corresponding group~\cite{huang2026hyformer}.

We focus on a less explored direction, in which each behavior reads non-sequential tokens (Figure~\ref{fig:maps}d,e). InterFormer conditions its sequence on a summary of non-sequential features that includes the candidate~\cite{zeng2024interformer}, so the conditioned sequence differs for every candidate and misses R2. We restrict the update to request tokens, built from the search query, user features and scene. This meets all three requirements. Each behavior sees the current search (R1). Because the request tokens are independent of the candidate, all candidates share the updated sequence (R2). With a fixed number of request tokens, attention cost grows linearly with $L$ (R3). Each behavior group also reads the same request representation while keeping its own ordering and projections, so heterogeneous groups need not be merged.

\textbf{\dna{} (Dual-stream Nexus Attention)} pairs this update with the usual cross-attention read, so attention runs in both directions at every layer. Both the request and candidate branches read the current behavior sequence through cross-attention. The sequence update then uses the new request representation as its keys and values, with each behavior token providing its own query (Figures~\ref{fig:maps}e and~\ref{fig:arch}). The next layer projects its keys and values from the updated behavior tokens. As the stack gets deeper, those tokens accumulate information from successive rounds of interaction with the request. The request branch reads across all behavior groups before each update, so information from a purchase in one group can reach a click in another through the updated request tokens. Each behavior keeps its own token throughout.

Leaving the candidate out of the update does not cost accuracy: in our experiments, the request-level design matches or exceeds the AUC of the candidate-aware design at the depths tested (Section~\ref{sec:who}). It also keeps serving cheap. At inference, the request branch, the sequence updates and the sequence key--value projections can all run once per request. Each candidate reads the shared keys and values with its own queries, and the prediction head combines the candidate representation with the shared request representation. The extra work is a fixed cost per request, while each candidate's reads cost the same as in a cross-attention stack of equal depth.

\begin{samepage}
\noindent\textbf{Our contributions.}\par\nopagebreak[4]
\begin{list}{\textbullet}{%
  \setlength{\leftmargin}{1.3em}%
  \setlength{\labelsep}{0.5em}%
  \setlength{\topsep}{3pt}%
  \setlength{\partopsep}{0pt}%
  \setlength{\itemsep}{2pt}%
  \setlength{\parsep}{0pt}}
\item \textbf{Problem.} We identify the fixed behavior sequence on the key--value side as the bottleneck that limits depth in cross-attention rankers, and show that letting the sequence evolve with depth sustains the gains from added layers (Sections~\ref{sec:space} and~\ref{sec:depth}).
\item \textbf{Method.} \dna{} couples sequence reads with request-conditioned sequence updates. It retains one token per behavior, connects multiple behavior groups through one request representation, and computes each update once per request at linear cost in sequence length (Section~\ref{sec:method}).
\item \textbf{Evidence.} On Taobao search, \dna{} leads the cross-attention stack by 0.08--0.26\pp{} AUC across six comparable forward-compute budgets and keeps improving with depth after the stack begins to level off. \dna[4] exceeds the 28-layer stack at about 40\% of its FLOPs (Section~\ref{sec:depth}).
\item \textbf{Deployment.} \dna[4] is deployed in Taobao search. Each sequence update runs once per request, so the estimated serving compute for a request with thousands of candidates is close to that of a cross-attention stack of the same depth. In an online A/B test against the production model, it improves CTR by 0.8\% and GMV by 0.5\% (Section~\ref{sec:deploy}).
\end{list}
\end{samepage}

\section{Problem Setup and Design Space}
\label{sec:space}

\textbf{Inputs and notation.} A request has user and request features, several groups of user behaviors, and $N$ candidate items. The behavior groups may differ in length, time span and feature fields. We project each group separately and retain its boundaries. This gives $\Sb_0\in\R^{L\times d}$, where $L=\sum_{g=1}^{G}L_g$ is the number of behavior positions, $G$ is the number of groups, and $d$ is the model width. We mask padded positions.

The group partition identifies which projection to apply to each behavior. A behavior belongs to its original group throughout the stack, and its position is retained. The concatenation therefore provides a common tensor interface for attention while preserving the groups' separate representations. The model does not need to sort events from different groups onto one timeline.

We use the HHFT feature tokenizer~\cite{yu2025hhft}, which maps semantic feature groups to $d$-dimensional tokens. It maps all non-sequential features, including those of the current candidate, to $\Fb_0\in\R^{m_F\times d}$. A tokenizer of the same form maps request-level features, such as the search query, user features and scene, to $\Ub_0\in\R^{m_U\times d}$. Every candidate has its own $\Fb_0$. All candidates of a request share $\Sb_0$ and $\Ub_0$. We use $\Sb_l$, $\Fb_l$ and $\Ub_l$ for these three representations before layer $l$. The model learns from request--candidate pairs with binary cross-entropy on click labels.

The two non-sequential arrays serve different roles. $\Fb_0$ can distinguish candidates for the same request, whereas $\Ub_0$ supplies a shared representation that can be used to update $\Sb$. Their token counts $m_F$ and $m_U$ are set independently. The model has no requirement to align individual tokens between these arrays.

\textbf{Baseline with a fixed sequence.} Let $\rblk_l^F$ denote the Read Block that updates $\Fb_l$ by attending to the sequence. A depth-$D$ cross-attention stack reads $\Sb_0$ at every layer,
\begin{equation}
\Fb_{l+1}=\rblk_l^F(\Fb_l,\Sb_0),
\qquad l=0,\ldots,D-1.
\label{eq:baseline}
\end{equation}
Each Read Block uses its own key and value projections of $\Sb_0$. STCA and MixFormer likewise transform the input sequence separately at each layer~\cite{guan2025stca,huang2026mixformer}. In this design the representation of one behavior can change through a position-wise projection, but it receives no information from other tokens. We call the $D$-layer baseline \ro[$D$]. Its two-layer version is the Taobao search control.

The distinction is between the state being read and its layer-specific projections. A Read Block can learn a different key and value map at every depth, but with fixed $\Sb_0$ its keys are always computed afresh from the original behaviors. The previous layer may have changed $\Fb$, yet no behavior has received any of that information. Our experiments therefore compare scaling the reads with scaling the sequence updates.

\textbf{Choices for updating $\Sb$.} Figure~\ref{fig:maps} places the baseline among the ways to update $\Sb$. Full self-attention (a) conditions the sequence on everything, but its cost is quadratic in $L$ and the sequence differs for every candidate. The \textbf{fixed-sequence baseline} (c) reads $\Sb_0$ at every layer. It can be shared across candidates (R2) and read at linear cost for a fixed number of queries (R3), but it never reflects the request. \textbf{Updating with $\Sb_l$} (b), as in HSTU, OneTrans and EST~\cite{zhai2024hstu,zhang2025onetrans,liu2026est}, keeps the sequence shareable, but the update draws only on the history, and full causal attention over $\Sb_l$ costs $O(L^2)$. \textbf{Updating with $\Fb_l$} (d), as in \dnat{}, makes the sequence request-conditioned (R1) at linear cost, since each behavior attends to a fixed number $m_F$ of tokens, but it produces a separate sequence for each candidate. In \dna{} (e), the request first reads $\Sb_l$, and the model then \textbf{updates with $\Ub_{l+1}$}. This meets all three requirements: the sequence incorporates the search query, all candidates share it, and each behavior attends to a fixed number $m_U$ of request tokens.

\section{The DNA Architecture}
\label{sec:method}

\begin{figure}[t]
\centering
\includegraphics[width=\columnwidth]{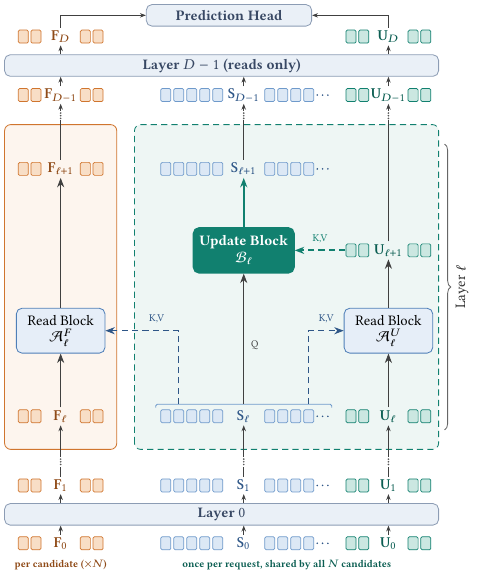}
\Description{The stack runs from the input tokens at the bottom to the prediction head at the top. In the expanded layer, both Read Blocks attend to the sequence, and the Update Block uses the updated request tokens to update the sequence. The final layer performs only the reads. Its candidate and request outputs enter the prediction head.}
\caption{\dna{} architecture, read from bottom to top, with layer $l$ shown in detail. The shaded regions separate the per-candidate $\Fb$ stream from the request-shared $\Sb$ and $\Ub$ streams. The final layer performs both reads without updating $\Sb$. The prediction head uses $\Fb_D$ and $\Ub_D$ to score each candidate. Removing the request stream $\Ub$ and the Update Blocks gives the baseline \ro[$D$].}
\label{fig:arch}
\end{figure}

\subsection{Overview}
Each \dna{} layer keeps the cross-attention read of the baseline and adds an update of the sequence by the request. Figure~\ref{fig:arch} follows the input states $\Fb_0$, $\Sb_0$, and $\Ub_0$ through the $D$-layer stack. In the expanded layer, the two Read Blocks, $\rblk_l^F$ and $\rblk_l^U$, attend to the same keys and values from $\Sb_l$ to produce $\Fb_{l+1}$ and $\Ub_{l+1}$. In the first $D-1$ layers, the Update Block $\wblk_l$ then applies cross-attention in the reverse direction: $\Sb_l$ supplies the queries and the new $\Ub_{l+1}$ supplies the keys and values. The updated sequence $\Sb_{l+1}$ is read by both branches in the next layer.

This exchange between $\Ub$ and $\Sb$ is the dual stream in \dna{}. Both states are shared within a request. Each candidate has its own $\Fb$, which reads the shared sequence independently.

\subsection{Read Block}
\label{sec:read}
The two Read Blocks form queries from $\Fb_l$ and $\Ub_l$ using separate learned projections of the normalized inputs.
\begin{align}
\Qb_l^F &= \operatorname{Proj}_{Q,l}^F\big(\mathrm{LN}(\Fb_l)\big),\\
\Qb_l^U &= \operatorname{Proj}_{Q,l}^U\big(\mathrm{LN}(\Ub_l)\big).
\end{align}
The query arrays have shapes $m_F\times d$ and $m_U\times d$, respectively. Both Read Blocks obtain keys and values from $\Sb_l$ using projections that depend on the group $g(j)$ of behavior token $j$.
\begin{equation}
\begin{aligned}
\Kb_{l,j}&=\mathbf{W}^{K}_{l,g(j)}\mathbf{s}_{l,j},\\
\Vb_{l,j}&=\mathbf{W}^{V}_{l,g(j)}\mathbf{s}_{l,j}.
\end{aligned}
\end{equation}
We compute these projections once per layer and share them between $\rblk_l^F$ and $\rblk_l^U$. The queries remain separate. The same sequence token can thus receive different attention weights from each candidate and from $\Ub$, even though these reads use identical keys and values.

For each head, let a tilde denote QKNorm applied to its queries or keys~\cite{henry2020qknorm}. In $\rblk_l^F$, attention weights are
\begin{equation}
\Pb_l^F=\mathrm{softmax}_{L}\!\left(
\frac{\widetilde{\Qb}_l^F\widetilde{\Kb}_l^\top}{\sqrt{d_h}}+\mathbf{b}_{\mathrm{pad}}
\right),
\end{equation}
where $d_h$ is the head width and $\mathbf{b}_{\mathrm{pad}}$ masks padded positions. For the Read Block on $\Fb$, we combine the attention heads and apply the output projection, followed by a token-specific SwiGLU FFN~\cite{shazeer2020glu}:
\begin{align}
\Fb'_l &= \Fb_l+\mathrm{ConcatHeads}(\Pb_l^F\Vb_l)\mathbf{W}^{F}_{l,O}, \notag\\
\Fb_{l+1} &= \Fb'_l+\mathrm{FFN}^{F}_l\big(\mathrm{LN}(\Fb'_l)\big).
\end{align}
The first residual adds information read from $\Sb_l$. The FFN applies a separate transformation to each token. The block $\rblk_l^U$ uses the same operations with $\Qb_l^U$ and $\Ub_l$, with its own output projection and FFN. The attention weights for the two reads have shapes $m_F\times L$ and $m_U\times L$ per head.

\subsection{Update Block}
The Update Block forms queries from the normalized sequence $\Sb_l$ using a separate projection for each behavior group. Two learned projections of the normalized request state $\Ub_{l+1}$ give the keys and values, each of shape $m_U\times d$. Following the composite projection of Hiformer~\cite{gui2023hiformer}, each is a single linear map applied to the concatenated request tokens, so every key and value can combine information from all request tokens.
\begin{align}
\Qb_{l,j}^{S} &= \mathbf{W}^{Q}_{l,g(j)}\,\mathrm{LN}(\mathbf{s}_{l,j}),\\
\Kb_l^U &= \operatorname{Proj}_{K,l}^U\big(\mathrm{LN}(\Ub_{l+1})\big),\\
\Vb_l^U &= \operatorname{Proj}_{V,l}^U\big(\mathrm{LN}(\Ub_{l+1})\big).
\end{align}
Each head assigns a separate distribution over the $m_U$ request tokens to every behavior:
\begin{equation}
\Pb_l^B=\mathrm{softmax}_{m_U}\!\left(
\frac{\widetilde{\Qb}_l^{S}(\widetilde{\Kb}_l^U)^\top}{\sqrt{d_h}}
\right).
\end{equation}
The attention weights have shape $L\times m_U$ per head. We apply them to $\Vb_l^U$, combine the heads, and update $\Sb_l$:
\begin{align}
\Sb'_l &= \Sb_l+\mathrm{ConcatHeads}(\Pb_l^B\Vb_l^U)\mathbf{W}^{B}_{l,O},\notag\\
\Sb_{l+1}&=\mathbf{1}_{\mathrm{valid}}\odot
    \big[\Sb'_l+\mathrm{FFN}^{B}_l\big(\mathrm{LN}(\Sb'_l)\big)\big].
\end{align}
Here $\mathrm{FFN}^{B}_l$ is a SwiGLU network with separate weights for each behavior group, and $\mathbf{1}_{\mathrm{valid}}$ zeros padded positions. The Update Block retains all behavior tokens. The attention weights determine how much information each behavior takes from each request token. Two behaviors in the same group share projection weights, but can choose different mixtures of request values. The update uses $\Ub_{l+1}$ rather than $\Ub_l$ because $\Ub_{l+1}$ has just read $\Sb_l$ across all groups, so each behavior receives what the request found in the whole history at this layer.

\subsection{Layer Schedule}
\label{sec:schedule}
Depth $D$ counts the Read Blocks in the $\Fb$ branch. At every layer, both Read Blocks read the incoming $\Sb_l$. For the first $D-1$ layers, the Update Block then reads the new $\Ub_{l+1}$.
\begin{align}
\Fb_{l+1}&=\rblk_l^F(\Fb_l,\Sb_l), && l=0,\ldots,D-1, \notag\\
\Ub_{l+1}&=\rblk_l^U(\Ub_l,\Sb_l), && l=0,\ldots,D-1, \notag\\
\Sb_{l+1}&=\wblk_l(\Sb_l,\Ub_{l+1}), && l=0,\ldots,D-2.
\label{eq:schedule}
\end{align}
The last layer has only the two Read Blocks, since no later layer would read an updated sequence. Its outputs, $\Fb_D$ and $\Ub_D$, are concatenated along the token axis and mean-pooled over all $m_F+m_U$ tokens:
\begin{equation}
\hat y=\sigma\!\left(\mathrm{MLP}\!\left(\mathrm{Mean}[\Fb_D;\Ub_D]\right)\right).
\end{equation}
This gives $\Ub$ a direct path to the click loss as well as a path through $\Sb$. In total, \dna[$D$] has $D$ Read Blocks for $\Fb$, $D$ for $\Ub$ and $D-1$ Update Blocks.

A two-layer model illustrates the order. The first pair of Read Blocks reads $\Sb_0$ and produces $\Fb_1$ and $\Ub_1$. The Update Block then builds $\Sb_1$ from $\Sb_0$ and $\Ub_1$. Both second-layer reads use $\Sb_1$. For two candidates of the same request, the first-layer $\Fb_1$ can differ, but $\Ub_1$ and $\Sb_1$ are shared. Each candidate's second-layer output is combined with the same $\Ub_2$ at prediction time.

\subsection{Request-Level Sequence Updates}
\label{sec:why}
\textbf{Request-level sharing.} Both $\Sb_0$ and $\Ub_0$ are candidate-independent. Equation~\eqref{eq:schedule} carries this property through the stack. We can therefore compute $\Ub$, update $\Sb$ and project its keys and values once per request. Only the $\Fb$ branch and prediction head run per candidate.

\textbf{Request-conditioned behavior representations.} $\Ub$ carries the current search query, user features and scene. After an update, a behavior token depends on its own features and on the request. The next Read Block uses this updated representation when it scores a candidate.

\textbf{Interaction across behaviors.} Each Read Block for $\Ub$ reads every group in $\Sb_l$. The updated $\Ub_{l+1}$ then passes information back to every group. Thus behavior $j$ can influence behavior $k$ through $\Ub_{l+1}$, using two attention operations with $O(m_U L)$ pairs in total. The groups keep separate projections and feature schemas without requiring a common event timeline. STCA likewise de-emphasizes explicit behavior-to-behavior attention~\cite{guan2025stca}.

\textbf{The candidate-aware variant.} In \dnat{}, $\Fb_l$ supplies the keys and values for the Update Block. The read is $\Fb_{l+1}=\rblk_l^F(\Fb_l,\Sb_l)$ and the update is $\Sb_{l+1}=\wblk_l(\Sb_l,\Fb_l)$. The former runs $D$ times and the latter $D-1$ times. The prediction uses $\Fb_D$. This variant tests whether letting the candidate into the update improves accuracy. It also makes $\Sb_l$ candidate-specific from the first update onward. Section~\ref{sec:who} compares the two choices.

\subsection{Complexity and Serving Cost}
\label{sec:cost}
A request with $N$ candidates costs $C_{\mathrm{req}}+N C_{\mathrm{cand}}$. Here $C_{\mathrm{req}}$ is the work shared within the request, and $C_{\mathrm{cand}}$ is the work for each candidate.

Projecting $\Sb_l$ to keys and values costs $c_{\mathrm{KV}}=O(Ld^2)$ per layer. Let $c_A(m)$ denote the remaining cost of a Read Block with $m$ queries. Attention costs $O(mLd)$, and the query and output projections and the FFN operate on the $m$ non-sequential tokens. Let $c_B(m)$ denote the cost of an Update Block with $m$ key--value tokens. Its attention also costs $O(mLd)$. The projections and FFN on the sequence cost $O(Ld^2+Ldh_S)$, where $h_S$ is the sequence FFN width. The key and value projections operate on the $m$ non-sequential tokens. With fixed $m_F$, $m_U$ and model widths, both block costs grow linearly with $L$.

\begin{itemize}
\item \textbf{\ro[$D$].} The request shares $D c_{\mathrm{KV}}$. Each candidate pays $D c_A(m_F)$.
\item \textbf{\dna[$D$].} Each candidate pays $D c_A(m_F)$, as in \ro[$D$]. The shared cost is $D c_{\mathrm{KV}}+D c_A(m_U)+(D-1)c_B(m_U)$.
\item \textbf{\dnat[$D$].} The request shares only the projection of $\Sb_0$, costing $c_{\mathrm{KV}}$. Each candidate pays for its reads and the updates and projections that follow, for a total of $D c_A(m_F)+(D-1)[c_{\mathrm{KV}}+c_B(m_F)]$.
\end{itemize}

DNA adds request-level work to a same-depth read stack. Because the updated $\Sb$ depends on the request, that work is repeated for new requests. The cost model counts FLOPs for the Read and Update Blocks. Table~\ref{tab:cost} evaluates it for our configuration in Section~\ref{sec:serving}.

Parameters and FLOPs grow at different rates in this design. A weight that acts on the sequence is applied at all $L$ positions, whereas a weight that acts on non-sequential tokens is applied to only $m_F$ or $m_U$ tokens. In \dna[4], operations over $\Sb$ account for about 85\% of the block FLOPs. We therefore compare models at similar forward FLOPs and report dense parameter counts separately.

\section{Experiments}
\label{sec:exp}

Three questions follow from the design. RQ1 asks whether updating $\Sb$ makes added depth pay off (Section~\ref{sec:depth}). RQ2 asks whether sequence updates need the target item, or whether request features suffice (Section~\ref{sec:who}). RQ3 asks what the design costs when a request scores thousands of candidates (Section~\ref{sec:serving}).

\subsection{Setup}
\textbf{Data.} We train on Taobao search impression logs and evaluate on the logs of the following day. The inputs include multiple user behavior sequences, with a combined padded length on the order of $10^3$ tokens. Sparse embeddings start from a common checkpoint. Dense parameters are trained from scratch.

\textbf{Metrics.} We report AUC changes in percentage points (pt), computed as $100(\mathrm{AUC}-\mathrm{AUC}_{\mathrm{ref}})$. Each table specifies its reference model. Dense parameter counts exclude embedding tables. Forward FLOPs count the main matrix multiplications, with two FLOPs per multiply--add. We report compute per sample for offline comparisons and per request for serving.

\textbf{Models.} All models share the HHFT feature tokenizer, a token-preparation layer~\cite{zhu2025rankmixer}, the base layers, the prediction head and the training recipe. The base layers include one Transformer and one Hiformer layer over non-sequential tokens~\cite{yu2025hhft}. \ro[$D$] uses $D$ Read Blocks for $\Fb$ and reads $\Sb_0$ throughout. \dna[$D$] follows Equation~\eqref{eq:schedule}. \dnat[$D$] uses $\Fb_l$ in the Update Block. The model width is 512 and each attention block has eight heads. All variants use pre-normalization and QKNorm.

\subsection{RQ1: Does Updating the History Make Depth Pay Off?}
\label{sec:depth}

\begin{figure*}[!t]
  \centering
  \includegraphics[width=\textwidth]{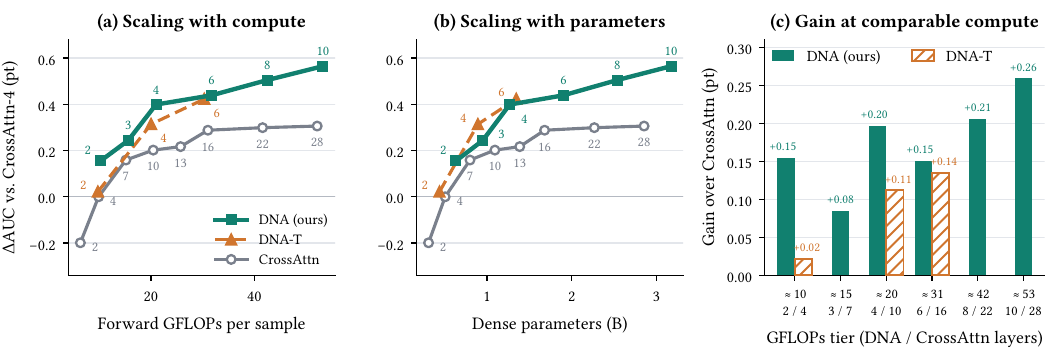}
  \Description{Three panels: AUC gains versus forward FLOPs and dense parameter count for CrossAttn, DNA and DNA-T, followed by AUC gaps relative to CrossAttn across six comparable compute tiers.}
  \caption{Scaling on Taobao search. (a) $\Delta$AUC over \ro[4] as a function of forward FLOPs per sample. (b) The same results as a function of dense parameter count. Numbers beside the markers in (a) and (b) denote model depth $D$. (c) AUC gaps relative to \ro{} at comparable forward compute, with the layer counts of \dna{} and \ro{} under each tier.}
  \label{fig:depth}
\end{figure*}

\begin{table}[!t]
\centering
\captionsetup{skip=5pt}
\caption{Accuracy at comparable forward compute. $\Delta$ is relative to \ro[4]. DNA gain is relative to \ro{} at the same tier. AUC differences are in pt.}
\label{tab:main}
\normalsize
\renewcommand{\arraystretch}{1.03}
\setlength{\tabcolsep}{2pt}
\newcommand{\depthdelta}[2]{\makebox[11pt][r]{#1}\hspace{4pt}\makebox[27pt][r]{#2}}
\begin{tabular*}{\columnwidth}{@{\extracolsep{\fill}}ccccr@{}}
\toprule
GFLOPs & \ro{} & \dnat{} & \dna{} & DNA gain\\
\cmidrule(lr){2-2}\cmidrule(lr){3-3}\cmidrule(lr){4-4}
per sample & \depthdelta{$D$}{$\Delta$} & \depthdelta{$D$}{$\Delta$} & \depthdelta{$D$}{$\Delta$} & (pt)\\
\midrule
$\approx$6 & \depthdelta{2}{$-$0.199} & \depthdelta{--}{--} & \depthdelta{--}{--} & --\\
$\approx$10 & \depthdelta{4}{0} & \depthdelta{2}{+0.022} & \depthdelta{2}{+0.155} & +0.155\\
$\approx$15 & \depthdelta{7}{+0.159} & \depthdelta{--}{--} & \depthdelta{3}{+0.244} & +0.084\\
$\approx$20 & \depthdelta{10}{+0.202} & \depthdelta{4}{+0.314} & \depthdelta{4}{+0.399} & +0.197\\
$\approx$26 & \depthdelta{13}{+0.217} & \depthdelta{--}{--} & \depthdelta{--}{--} & --\\
$\approx$31 & \depthdelta{16}{+0.288} & \depthdelta{6}{+0.423} & \depthdelta{6}{+0.439} & +0.152\\
$\approx$42 & \depthdelta{22}{+0.299} & \depthdelta{--}{--} & \depthdelta{8}{+0.505} & +0.205\\
$\approx$53 & \depthdelta{28}{+0.306} & \depthdelta{--}{--} & \depthdelta{10}{\textbf{+0.565}} & \textbf{+0.259}\\
\bottomrule
\end{tabular*}
\end{table}

We match models by single-sample forward FLOPs. A \dna{} layer costs about as much as two to three layers of \ro{}, so the comparable \ro{} stack is deeper. Figure~\ref{fig:depth} plots the same runs against forward FLOPs (a) and dense parameters (b), and shows the AUC gap at each comparable compute tier (c). Table~\ref{tab:main} lists the six paired tiers, along with \ro[2] and \ro[13] as unpaired points on the baseline curve.

\textbf{The fixed-sequence baseline levels off.} Adding layers to \ro{} improves AUC by 0.20\pp{} from 4 to 10 layers and by 0.09\pp{} from 10 to 16. Growing the stack from 16 to 28 layers increases forward FLOPs by 68\% for a further 0.02\pp{}.

\textbf{Updating $\Sb$ improves the returns to depth.} \dna{} leads \ro{} by 0.08 to 0.26\pp{} across all six paired budgets. \dna[2], with a single Update Block, leads \ro[4] by 0.15\pp. \dna[4] already exceeds \ro[28] by 0.09\pp{} at about 40\% of its FLOPs. The gap widens from 0.15 to 0.26\pp{} between roughly 31 and 53 GFLOPs, where the baseline levels off. Over this compute increase, \ro{} grows from 16 to 28 layers and gains 0.02\pp{}, while \dna{} grows from 6 to 10 layers and gains 0.13\pp{} (Figure~\ref{fig:teaser}).

\textbf{Comparing at equal depth.} Matching FLOPs asks which architecture makes better use of a per-sample compute budget. Matching depth instead holds the per-candidate work fixed, because both stacks then have the same Read Blocks in the $\Fb$ branch (Section~\ref{sec:cost}). At equal depth, \dna[2], \dna[4] and \dna[10] lead their \ro{} counterparts by 0.35, 0.40 and 0.36\pp{}. Section~\ref{sec:serving} accounts for the shared work that \dna{} adds.

\textbf{Parameter efficiency.} Figure~\ref{fig:depth}(b) also shows gains at similar or lower dense parameter counts. \dna[8] exceeds \ro[28] by 0.20\pp{} with 18.5\% fewer FLOPs and 11\% fewer parameters. At the 20-GFLOP tier, \dna[4] has 16\% more dense parameters than \ro[10] (1.267B versus 1.093B); Section~\ref{sec:who} evaluates a reduced \dna[4] with fewer parameters.

\subsection{RQ2: Does the Update Need the Target Item?}
\label{sec:who}

\begin{table}[!tb]
\caption{Sequence-update ablations at $D=4$. AUC changes are in pt relative to the first row of each group. (a) Update inputs and dense parameter counts (billions). $\Fb$ denotes all non-sequential tokens and $\Ub$ denotes request tokens. (b) Components of the \dnat[4] Update Block.}
\label{tab:who}
\centering
\normalsize
\renewcommand{\arraystretch}{1.13}
\begin{tabularx}{\columnwidth}{@{}Xcrr@{}}
\toprule
Model & Update input & Dense (B) & $\Delta$AUC\\
\midrule
\multicolumn{4}{@{}l}{\emph{(a) Source of the update}}\\
\dnat[4] & $\Fb$ & 0.892 & 0\\
\dna[4] & $\Ub$ & 1.267 & +0.09\\
\dna[4] (reduced) & $\Ub$ & 0.937 & +0.01\\
\midrule
\multicolumn{4}{@{}l}{\emph{(b) Update block}}\\
\multicolumn{3}{@{}l}{Cross-attention to $\Fb$ and FFN} & 0\\
\multicolumn{3}{@{}l}{\quad attention only} & $-$0.037\\
\multicolumn{3}{@{}l}{\quad FFN only} & $-$0.303\\
\bottomrule
\end{tabularx}
\end{table}

\textbf{Request features suffice for sequence updates.} \dnat{} improves on \ro{} at all three budgets where we test it, by up to 0.14\pp{}. With request-only sequence updates, \dna{} exceeds \dnat{} by 0.13, 0.09 and 0.02\pp{} at depths 2, 4 and 6, respectively (Table~\ref{tab:main}).

Table~\ref{tab:who}(a) compares three four-layer models. \dnat[4] updates the sequence from $\Fb_l$, which includes candidate features. \dna[4] adds a Read Block for $\Ub$ at each layer and updates the sequence from $\Ub_{l+1}$. The full model has more parameters than \dnat[4], so its 0.09\pp{} lead does not isolate the effect of changing the update input. Replacing the composite key and value projections of its Update Blocks with standard per-token projections reduces the dense parameter count of \dna[4] from 1.267B to 0.937B, close to the 0.892B of \dnat[4]. The reduced model is just 0.01\pp{} above \dnat[4]. At similar parameter counts, request-only updates therefore retain the accuracy of candidate-aware updates while allowing all candidates to share the updated sequence.

\textbf{Cross-attention drives the update.} Removing cross-attention from the \dnat[4] Update Block and retaining only its FFN loses 0.30\pp{}, whereas removing the FFN loses 0.037\pp{} (Table~\ref{tab:who}(b)). The per-behavior FFN transforms each token independently. Cross-attention lets it incorporate information from $\Fb_l$. The much larger loss from removing cross-attention shows that cross-side interaction is the main contributor to the update gains in this comparison.

At depth four, using $\Ub$ to update $\Sb$ also reduces the per-candidate compute from 15.7 to 0.94 GFLOPs under the block cost model (Table~\ref{tab:cost}).

\subsection{RQ3: Serving Cost}
\label{sec:serving}

\begin{figure}[!t]
  \centering
  \includegraphics[width=\columnwidth]{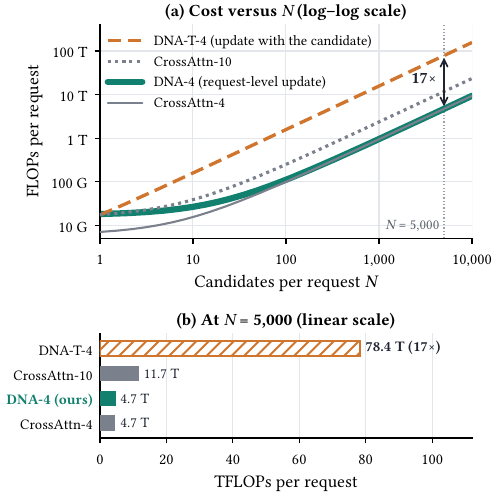}
  \Description{Two panels. Top: log-log plot of the compute of one request against the number of candidates for DNA-4, the target-aware variant and two cross-attention stacks. Bottom: the same costs at 5,000 candidates on a linear axis, where the target-aware variant costs 78.4 TFLOPs and DNA-4 4.7 TFLOPs.}
  \caption{Estimated block-level forward FLOPs per request. (a) Cost versus the number of candidates $N$ on log--log axes. \dna[4] updates the sequence once per request and adds the same cost per candidate as \ro[4], so their curves nearly overlap at large $N$. (b) The same costs at $N=5{,}000$ on a linear axis. \dnat[4] repeats the sequence updates and the resulting K/V projections for each candidate and costs about $17\times$ as much as \dna[4].}
  \label{fig:serving}
\end{figure}

Single-sample FLOPs count one pass over the sequence and one candidate branch. That is a useful comparison for the offline experiment, but requests can contain many candidates. Once a request contains several items, the cost depends on which parts of that pass have to be repeated. The two quantities in Section~\ref{sec:cost} give each curve in Figure~\ref{fig:serving}(a). The shared cost determines its intercept, and the per-candidate cost determines the increase as each candidate is added.

\begin{table}[!t]
\caption{Block-level FLOPs by model. G and T denote GFLOPs and TFLOPs. Per-sample cost is $C_{\mathrm{req}}+C_{\mathrm{cand}}$; request totals use $N=5{,}000$ candidates. The tokenizer, base layers and prediction head are excluded (about 2.8 GFLOPs per sample in total).}
\label{tab:cost}
\centering
\normalsize
\renewcommand{\arraystretch}{1.2}
\setlength{\tabcolsep}{3pt}
\begin{tabular*}{\columnwidth}{@{\extracolsep{\fill}}lrrrr@{}}
\toprule
Model & \shortstack{Shared\\(G)} & \shortstack{Per cand.\\(G)} & \shortstack{Per sample\\(G)} & \shortstack{Request\\(T)}\\
\midrule
\ro[2]  & 3.1  & 0.47  & 3.5  & 2.34\\
\ro[4]  & 6.1  & 0.94  & 7.1  & 4.69\\
\ro[10] & 15.3 & 2.34  & 17.6 & 11.72\\
\ro[28] & 42.8 & 6.55  & 49.4 & 32.80\\
\dnat[4]& 1.5  & 15.68 & 17.2 & 78.41\\
\dna[4] & 17.2 & 0.94  & 18.1 & 4.70\\
\dna[8] & 37.8 & 1.87  & 39.7 & 9.40\\
\dna[10]& 48.1 & 2.34  & 50.4 & 11.75\\
\bottomrule
\end{tabular*}
\end{table}

\textbf{Request-level work is amortized across candidates.} Table~\ref{tab:cost} and Figure~\ref{fig:serving} show how the comparison changes when a request scores thousands of candidates. About 95\% of the block FLOPs of \dna[4] are shared at request level. At $N=5{,}000$, \dna[4] costs 4.70 TFLOPs. This is within 1\% of \ro[4], which it exceeds by 0.40\pp{} AUC, and about 40\% of the cost of \ro[10], which has similar single-sample FLOPs.

At the same candidate count, \dna[10] costs 11.75 TFLOPs. It costs about as much as \ro[10] and 36\% as much as \ro[28], and it exceeds their AUC by 0.36 and 0.26\pp{}, respectively. \dnat[4] repeats its sequence updates and most sequence projections for each candidate. It costs about 78 TFLOPs per request, roughly 17 times as much as \dna[4] at lower AUC (Figure~\ref{fig:serving}b).

The curves distinguish two ways of spending extra compute. At $N=1$, \dna[4] is close to \ro[10] because it runs both Read Blocks and sequence updates. At larger $N$, \ro[10] repeats ten reads for every candidate. \dna[4] repeats four and uses the same shared sequence state for each of them. Comparing \dna[4] with \ro[4] holds these four reads fixed. The former adds about 11 GFLOPs of shared work per request, so their relative cost gap shrinks as $N$ grows.

The \dnat[4] comparison changes the slope. Its first $\Sb_0$ projection is still shared. All three subsequent updates and the projections of their outputs depend on which candidate supplied $\Fb_l$. The cost of this work therefore grows with $N$. This difference follows from the update input even though the two four-layer models have comparable forward FLOPs per sample.

\section{Online Deployment}
\label{sec:deploy}

\textbf{Serving.} The request stage tokenizes $\Ub_0$, runs the $D$ Read Blocks for $\Ub$ and the $D-1$ Update Blocks, and projects each $\Sb_l$ to keys and values. The candidate stage runs the $D$ Read Blocks for $\Fb$ and the prediction head. All candidates read one copy of the keys and values at each layer. The $Nm_F$ queries for that layer can be combined in one batched attention call~\cite{dao2022flashattention,juravsky2024hydragen}.

The stages can be evaluated separately because their dependencies differ. The request stage obtains $\Ub_{l+1}$ directly from $\Sb_l$ and can compute $\Sb_{l+1}$ immediately. When the candidates are scored, their Read Blocks consume the saved sequence projections. Their attention weights and $\Fb$ outputs still vary by candidate. The head uses the individual $\Fb_D$ together with the shared $\Ub_D$.

This reuse applies within a request. A new search query can produce a different $\Ub_0$ even when the user history has not changed. The update can then produce different sequence representations. The FLOP estimates in Section~\ref{sec:serving} count the shared stage for each request, rather than assuming that the updated states can be carried across requests.

\textbf{Online A/B test.} We deployed \dna[4] in Taobao search and used \ro[2], the production model, as the control. In the offline comparison, \dna[4] is 0.60\pp{} above \ro[2] and 0.40\pp{} above \ro[4], which has the same per-candidate block cost. In the online A/B test, \dna[4] improves CTR by 0.8\% and GMV by 0.5\%.

\section{Related Work}
\label{sec:related}

\textbf{Reading long behavior sequences.} DIN introduces target attention over the behavior history~\cite{zhou2018din}. SIM and TWIN retrieve a subset of behaviors from lifelong histories before applying attention~\cite{pi2020sim,chang2023twin}. BST and SASRec encode behaviors with self-attention~\cite{chen2019bst,kang2018sasrec}, while TransAct incorporates the candidate into its sequence encoder~\cite{xia2023transact}. STCA and MixFormer use a stack of cross-attention layers. Each layer projects the original behavior tokens again, so the sequence inputs stay fixed across the stack~\cite{guan2025stca,huang2026mixformer}. LONGER reads the full sequence once and then applies self-attention among a smaller set of tokens~\cite{chai2025longer}.

\textbf{Updating behavior representations.} HSTU and OneTrans use causal attention across behavior tokens~\cite{zhai2024hstu,zhang2025onetrans}. OneTrans also prunes sequence tokens as depth increases. OneTrans-V2 encodes append-only history as candidate-independent shared context~\cite{cao2026onetransv2}. EST aggregates content-similar behaviors within each sequence, and MARM reuses target-attention outputs cached from earlier requests~\cite{liu2026est,lv2024marm}. These methods can change behavior representations with depth. HyFormer decodes each sequence with its own queries and supports several sequence encoders, including self-attention, LONGER-style cross-attention and a position-wise SwiGLU map~\cite{huang2026hyformer}.

InterFormer summarizes non-sequential features, including the candidate, to condition its sequence transformation~\cite{zeng2024interformer}. It applies this transformation across behaviors and uses self-attention over the merged sequence. In \dna{}, each behavior attends to $\Ub$ and can use a different mixture of request tokens. This keeps the update independent of the candidate and avoids full attention over $\Sb$. The same $\Ub$ connects behavior groups with different schemas.

\textbf{Sharing work within a request.} HSTU uses M-FALCON, and STCA batches candidates from a request~\cite{zhai2024hstu,guan2025stca}. LONGER, OneTrans and PinFM reuse sequence keys and values~\cite{chai2025longer,zhang2025onetrans,chen2025pinfm}. MixFormer, UG-Separation, ROO and ROCS also design for request-level reuse~\cite{huang2026mixformer,lu2026ugsep,guo2025roo,chen2026rocs}. \dna{} extends reuse to sequence updates by keeping $\Sb$ candidate-independent throughout the stack.

\textbf{Scaling feature interactions.} DLRM, DCNv2, AutoInt, Hiformer and HHFT model interactions among non-sequential features~\cite{naumov2019dlrm,wang2021dcnv2,song2019autoint,gui2023hiformer,yu2025hhft}. Wukong and RankMixer scale this part of the ranker~\cite{zhang2024wukong,zhu2025rankmixer}. Khrylchenko et al.\ scale sequential recommender transformers to one billion parameters~\cite{khrylchenko2025scaling}. Our experiments ask how depth changes the interaction between the non-sequential features and the behavior sequence.

\textbf{Bidirectional attention between short and long sequences.} ISAB, Luna, BiXT and RIN exchange information between a long input and a short set at linear attention cost~\cite{lee2019settransformer,ma2021luna,hiller2024bixt,jabri2023rin}. Perceiver encodes a long input into a latent set~\cite{jaegle2021perceiver,jaegle2022perceiverio}. \dna{} shares their linear two-way cross-attention, but its short stream is not a set of learned latents introduced for compression. It is the ranking context itself, and only its candidate-free part, $\Ub$, updates the sequence, so every candidate of a request can share the result.

\section{Conclusion}
Cross-attention stacks read the same behavior sequence at every layer, and on Taobao search their gains taper off after 16 layers. \dna{} lets the request read the sequence and then update it between layers, so added depth keeps paying off: at comparable forward FLOPs, \dna[10] exceeds a 28-layer cross-attention stack by 0.26\pp{} AUC.

The update uses only $\Ub$, so $\Sb$ remains shared across all candidates in a request. In our comparisons, this choice preserves accuracy relative to a candidate-aware update. At the same depth, it gives \dna{} the same per-candidate block cost as the cross-attention stack. The deployed \dna[4] improves CTR by 0.8\% and GMV by 0.5\% in the online A/B test. We plan to extend this approach to greater depth, longer histories and generative ranking.

\bibliographystyle{ACM-Reference-Format}
\bibliography{refs}

\end{document}